\documentclass[12pt]{iopart}
\usepackage{graphicx}

\usepackage{makecell}

\usepackage{graphicx}
\usepackage{tikz}
\usepackage{pgfplots}
\pgfplotsset{compat=1.5}
\usepackage{pgfplotstable}
\usetikzlibrary{calc}

\begin{document}

\title[]{Ultrafast Polarization-Tunable Monochromatic Extreme Ultraviolet Source at High-Repetition-Rate}

\author{Antoine Comby, Debobrata Rajak, Dominique Descamps, Stéphane Petit, Valérie Blanchet, Yann Mairesse, Jérome Gaudin, and Samuel Beaulieu}

\address{Universit\'e de Bordeaux - CNRS - CEA, CELIA, UMR5107, F33405 Talence, France}
\ead{samuel.beaulieu@u-bordeaux.fr}
\vspace{10pt}

\begin{abstract}
We report on the development of a high-order harmonic generation (HHG)-based ultrafast high-repetition-rate (250~kHz) monochromatic extreme ultraviolet (21.6~eV) source with polarization tunability, specifically designed for multi-modal dichroism in time- and angle-resolved photoemission spectroscopy. Driving HHG using an annular beam allows us to spatially separate the high-power 515~nm driving laser from the XUV beamlet while preserving the linear polarization axis angle tunability. This enables controlling the polarization state of the XUV radiation with a fixed all-reflective phase-shifter. This scheme leads to a source brightness of 5.7$\times$10$^{12}$ photons/s at 21.6~eV (up to 4$\times$10$^{11}$ photons/s on target) with ellipticities as high as 90$\%$. 
\end{abstract}

\section{Introduction}\label{sect_intro}

Photoemission spectroscopy (PES) is a powerful and universal technique to probe the electronic structure of matter~\cite{Reinert05}. PES is based on the photoelectric effect, in which bound electrons absorb one photon (typically in the ultraviolet -- UV, extreme ultraviolet -- XUV, or x-ray spectral range) with energy  larger than their binding energy and escape into vacuum~\cite{Hertz87,Einstein05}. Energy conservation dictates that energy levels of the sample are imprinted onto the kinetic energy spectrum of outgoing photoelectrons. In addition, by investigating more subtle properties of outgoing electrons (e.g. angular distribution, spin polarization, temporal profile, light polarization dependence), rich information about both the electronic structure of the irradiated sample and the photoemission process itself can be extracted. 

In atomic and molecular physics, the recent improvements of light sources and photoelectron detectors, combined with clever measurement protocols, have enabled impressive experimental achievements such as the measurement of molecular frame photoelectron angular distributions (MFPAD)~\cite{Dill76,Lafosse03,Bisgaard09}, the realization of so-called 'complete experiments', in which amplitudes and phases of complex transition dipole matrix elements are determined~\cite{Duncanson76,Reid03,Cherepkov_05,hockett_14,Villeneuve_17,Marceau17}, the measurement of attosecond delays in photoemission ~\cite{Wigner55,Haessler09,Schultze10,Kheifets10,Klunder11,beaulieu17,Isinger17}, and the investigation of the chirality of matter using photoionization with circularly polarized light~\cite{ritchie76,powis00,bowering01,nahon15,Ferre15}, to list only few examples. 

In condensed matter physics, angle-resolved photoemission spectroscopy (ARPES) is well established as a powerful technique to study the electronic structure of crystalline solids in their ground state~\cite{Damascelli04,Gedik17,Lv19,Sobota21}. Indeed, when photoemission occurs from solids, in addition to the energy conservation law, the conservation of in-plane momentum provides additional information about electron's momenta inside the crystal. Using high photon energy provides a photoemission horizon (defined as the accessible range in reciprocal space) reaching high parallel momenta. Thus, using a photon energy of 20~eV (in the XUV spectral range) provides a photoemission horizon larger than solid's entire Brillouin zone ($\sim \pm$ 1.5~$\mathrm{\AA}$), while using low photon energy sources, e.g. 6~eV, only gives access to small parallel momenta, i.e. close to the Brillouin zone center ($\sim \pm$ 0.5~$\mathrm{\AA}$). XUV-ARPES is therefore the workhorse of experimental condensed matter physicists to unveil the electronic band structure of solids within their entire Brillouin zone.

In the past decades, by scanning new experimental parameters and by resolving more degrees of freedom of the outgoing photoelectrons, new flavors of ARPES have emerged. For example, combining ARPES with a pump-probe scheme, time-resolved (tr)ARPES allows following the out-of-equilibrium dynamics of solids upon impulsive optical excitation~\cite{Haight85,Smallwood16}. By controlling geometric aspects of the light-matter interaction, e.g. light polarization state and sample orientation, many types of dichroic observables can be extracted: linear dichroism in the photoelectron angular distribution (LDAD) to extract the symmetry of the wavefunctions~\cite{Sterzi18,Volckaert19,Rostami19,Min19,Moser22} and circular dichroism in the photoelectron angular distribution (CDAD) to interrogate the topological and quantum-geometrical character of bands~\cite{Cho18,Cho21,Schuler20-1}. More recently, intrinsic linear~\cite{Beaulieu21-3}, time-reversal~\cite{beaulieu20-2} and Fourier dichroism~\cite{Beaulieu22} in the photoelectron angular distribution (\textit{i}LDAD, TRDAD and FDAD, respectively) were introduced, allowing to probe the orbital texture and reconstruct the Bloch wavefunction of solids. Combining these dichroic observables with the time-resolution offered by trARPES would allow following in real-time the ultrafast evolution of the local Berry curvature, orbital pseudospin, and topology of the electronic band structure undergoing dynamics in photoexcited solids~\cite{Schuler20-5,Schuler22-2}. 

While combining trARPES with all these dichroic observables is extremely appealing, it also comes with very challenging light source requirements. Indeed, such light source must simultaneously be: \textbf{(i)} bright and at high-repetition-rate -- for high signal-to-noise ratio, \textbf{(ii)} of ultrashort duration -- for time-resolution, \textbf{(iii)} in the XUV spectral range -- to access the entire Brillouin zone of solids, \textbf{(iv)} with a restricted spectral bandwidth -- to avoid overlapping photoelectron spectrum replica and last, but not least \textbf{(v)} polarization-tunable -- to generate differential dichroic observables described above. While some of these characteristics are individually reached in many HHG-based trARPES setups (e.g. see~\cite{Haight88,Frietsch13,Eich14,Cilento16,Rohde16,Corder18,Sie19,Buss19,Puppin19,Mills19,Lee20,Cucini20,Guo22}), we are not aware of any light source delivering all these performance simultaneously. This represents a challenge that the current work aims at overcoming.

In this article, we describe a light source for next-generation time-, angle- and polarization-resolved ARPES, combining simultaneously all these characteristics: bright (5.7$\times$10$^{12}$ photons/s at 21.6~eV -- 4$\times$10$^{11}$ photons/s on target), high-repetition-rate (250~kHz), ultrashort ($\sim$120~fs), monochromatic extreme ultraviolet photon energy (21.6~eV) as well as polarization state tunable (full control of the linear polarization axis direction, and ellipticity up to $\sim$90$\%$). 

The article is organized as follows: in section~\ref{sect_setup}, we provide a brief and general overview of the XUV beamline. In sections~\ref{sect_annular} to \ref{sect_polar}, we describe the choice of the different building blocks of our source based on a detailed analysis of the existing state-of-the-art XUV beamlines. In section~\ref{sect_annular}, we describe different techniques to remove the driving laser in HHG beamline, with emphasis on the annular beam approach allowing us to spatially separate the high-power 515~nm driving laser from the XUV beam while preserving the linear polarization axis angle control. In section~\ref{sect_scaling}, we describe the XUV flux optimization. In section~\ref{Fig_mono}, we review different techniques to spectrally isolate a single harmonic from the harmonic comb, focusing on our approach based on a combination of a multilayer mirror and a free-standing metallic foil. In section~\ref{Fig_polar}, we discuss the different approaches to generate circularly-polarized femtosecond XUV pulses and show how we manipulate the XUV polarization state using a home-built all-reflective XUV quarter-wave plate. In section~\ref{sect_stab}, we demonstrate the long-term stability of our circularly polarized XUV source by monitoring the electron count rate in single-photon ionization of gas-phase molecules over 40 hours. In section~\ref{sect_conclu}, we conclude by comparing the performance of our beamline with the already existing quasi-circularly polarized HHG-based XUV sources.

\section{General description of the XUV source}\label{sect_setup}

\begin{figure}[h]
\begin{center}
\includegraphics[width=1\textwidth, keepaspectratio=true]{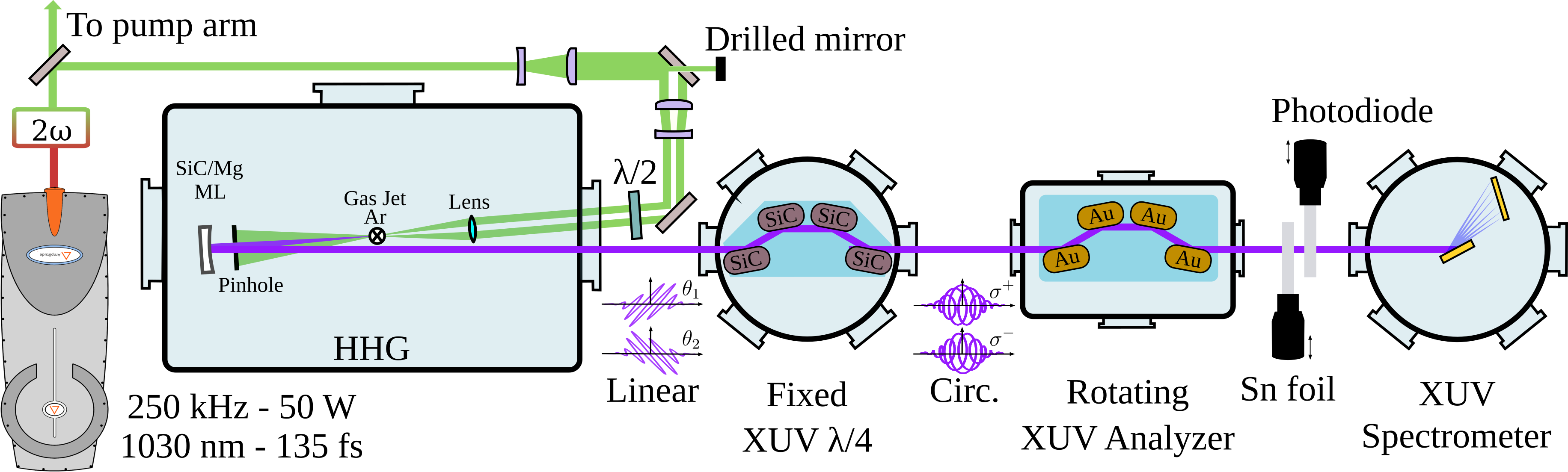}
\caption{\textbf{Scheme of the experimental setup}. The beamline features a frequency-doubled high-repetition-rate Yb fiber laser used to drive HHG in an annular beam configuration. Harmonic 9 (21.6~eV) is spectrally isolated using a combination of an XUV SiC/Mg multilayer mirror (ML) and a Sn metallic foil. The polarization state of the XUV beam is controlled by rotating the linear polarization axis angle impinging on a fixed all-reflective XUV quarter-wave plate. The polarization state of the 21.6~eV pulses is characterized by recording Malus law -- by rotating an XUV analyzer (reflection on four bare Au mirrors) in front of an XUV photon detector (more details about the setup can be found in the main text).}
\label{Fig_setup}
\end{center}
\end{figure}

A general overview of the experimental setup is given in Fig. \ref{Fig_setup}. Our setup is based on the use of the BlastBeat femtosecond laser system, at the Centre Lasers Intenses et Applications - CELIA (Bordeaux, France), which consists of two Yb-doped fiber lasers (seeded by the same oscillator) delivering two synchronized 50 W beams of 135~fs pulses centered around 1030~nm (spectral bandwidth FWHM = 18.5~nm), at a repetition rate which can be continuously tuned between 166~kHz and 2~MHz (Tangerine Short Pulse, Amplitude Group). Here, we operate the laser system at 250~kHz, which is a good trade-off to ensure good statistics in future photoemission experiments while maintaining enough energy per pulse to drive highly non-linear processes such as HHG. 

The fundamental laser beam (50~W, 1.3~mm radius at $I_{max}/e^2$)  is frequency-doubled in a 1~mm thick $\beta$-barium borate (BBO) crystal ($\theta$ = 23.4$^{\circ}$, $\phi$ = 90$^{\circ}$) to obtain a 515 nm average power up to 19 W (38$\%$ conversion efficiency, $\sim$0.9~mm radius at $I_{max}/e^2$, duration 130~fs FWHM)~\cite{Comby19,Comby20}. The BBO crystal is heated at a temperature of $\sim$100$^{\circ}$ to avoid long-term degradation (fogging) due to its hygroscopic nature. We use up to $\sim$35$\%$ of the 515~nm power to drive HHG, to keep most of the power available for frequency conversion in the pump arm in future pump-probe experiments. The 515~nm beam (up to 7~W) is expanded by a -75~mm/+300~mm ($\times$4) telescope before being reflected on a drilled mirror ($\phi$ = 4~mm) and reduced by a +125~mm/-75~mm ($\times$0.6) telescope. The two telescopes placed respectively before and after the drilled mirror allow us to finely adjust the size of the inner and outer part of the annular beam. The annular beam propagates through a motorized zero-order half-wave plate which enables controlling its linear polarization axis direction. The pulses are focused by an f=+79~mm aspheric lens (down to a focus size of $\sim$ 5x5$\mu$m FWHM) into a thin and dense argon gas jet (nozzle throat size 50$\mu$m) located inside a vacuum chamber. A movable (x,y,z) gas catcher (not shown in Fig.\ref{Fig_setup}) is placed a few mm above the exit of the gas jet to limit the XUV reabsorption from the residual gas. 

A pinhole placed $\sim$ 190~mm after the gas jet  blocks the 515~nm beam, while the XUV beam is collimated by an f=+200~mm focusing SiC/Mg multilayer mirror (NTTAT) located $\sim$ 200~mm away from the HHG source (under 5$^{\circ}$ angle of incidence). The reflectivity of the SiC/Mg multilayer spherical mirror is maximal (40$\%$) around 21.6~eV (harmonic 9) and partially suppresses the adjacent harmonics. Additional spectral selection is achieved by propagating the XUV beam through a free-standing 400~nm or 200~nm thick Sn metallic foil (5$\%$ and 22$\%$ transmission at 21.6~eV, respectively). The XUV beam then impinges onto a fixed all-reflective XUV quarter-wave plate made of four two inches SiC mirrors (NTTAT) under 78$^\circ$ angle of incidence. The XUV quarter-wave plate was designed to ensure a phase retardance between the s- and p- components of $\lambda$/4 at 21.6~eV, while maintaining a relatively good overall transmission of 41$\%$ in s-polarization and 21$\%$ in p-polarization. The polarization state can be characterized by recording Malus law by rotating an XUV analyzer before a photon detector made of a set of chevron microchannel plates (MCP) with a fast (P46) phosphor screen, which is imaged by a CCD camera. 

\section{Isolating the XUV beam from its driver}\label{sect_annular}

Most applications of HHG sources require getting rid of the driving laser beam after high-order harmonic generation -- the presence of the intense driving pulses in the XUV target region is generally undesired. To solve this issue, various approaches have been developed within the past decades, each having specific advantages and disadvantages. We review the most common approaches in Table~\ref{table1}.

The first straightforward approach is to take advantage of huge spectral separation of the driver and the harmonics and use thin metallic filters which reflect/absorb the driver while partially transmitting the XUV~\cite{Schins96}. The transmission depends on the chemical composition and thickness of the filter, and a variety of choices are available depending on the desired spectral range. However, since freestanding thin metallic filters absorb light and have a very low damage threshold, they easily melt under irradiation, which strongly limits the use of this approach for getting rid of high-power, high-repetition-rate driving lasers. These filters thus have to be placed after other elements which reduce the driving laser power. This can be achieved by propagation through microchannel plates, which strongly diffract the fundamental beam while keeping most XUV photons on axis~\cite{Zhang14}. This technique can thus be used to reduce the amount of laser power on a metal filter, but it degrades the spatial quality of the XUV beam.

Another appealing approach to separate the XUV and fundamental beams is to use beam splitting plates. For instance, a fused silica plate can be covered by an anti-reflective coating at the laser wavelength topped by an XUV reflecting layer (e.g. SiO$_2$, Nb$_2$O$_5$). The XUV radiation only sees top layers, and is properly reflected if the incidence angle is sufficiently grazing, while the laser beam is transmitted~\cite{Wabnitz06,Loch11}. The transmission of the laser through the plate can be further optimized by using Brewster incidence for the laser wavelength~\cite{Takahashi04,Pronin11}. Indeed, at the Brewster angle, the p-polarized driver (typically visible - infrared) is not reflected while the reflectivity in the XUV spectral range is typically in the tens of percent range. These techniques are routinely used on HHG beamlines, but rely on grazing incidence optics, whose reflectivity strongly depends on the polarization state of the XUV radiation. Consequently, they cannot be employed for experiments in which the polarization state is tuned. 

Gratings can also be used to separate the XUV and fundamental beams unique with concomitant advantage of XUV monochromatization properties~\cite{Mills12}.  However, single-grating configuration introduces spatial and temporal chirp. This can be circumvented by using double-grating monochromators~\cite{Poletto07}, whose efficiency can be optimized relying on conical diffraction. These grating-based techniques are generally costly and require very accurate alignment. Moreover, in our case, we cannot use diffraction gratings since their reflectivity strongly depends on the polarization state of the XUV radiation.

\begin{center}
\begin{table}
 \footnotesize
\begin{tabular}{||c | c | c | c | c | c||} 
 \hline
 \textbf{Methods} & \makecell{\textbf{Typical} \\ \textbf{Laser} \\ \textbf{Rejection}} & \makecell{\textbf{Typical} \\ \textbf{XUV} \\ \textbf{Trans.}} & \textbf{Pros} & \textbf{Cons} & \textbf{Ref.} \\ 
 \hline
 \makecell{Metallic \\ filter} & 100$\%$ & 5-50$\%$ & \makecell{Preserves polarization \\ Partial spectral selection}  & \makecell{Melts at high power} & \cite{Schins96} \\ 
 \hline
 \makecell{Diffractive \\ MCPs} & $>$90$\%$  & $\sim$25$\%$ & \makecell{Preserves polarization} & \makecell{Degrade the \\ spatial profile} & \cite{Zhang14} \\ 
 \hline
 \makecell{Single grating \\ monochromator} & 100$\%$ & $\sim$10$\%$ & Spectral selection & \makecell{Spatial and temporal \\ chirp \\ Polarization sensitive} & \cite{Mills12} \\
 \hline
 \makecell{Time-compensating \\ monochromator} & 100$\%$ & 2-20$\%$ & XUV spectral selection & Polarization sensitive & \cite{Poletto07,Frassetto11}\\
 \hline
 \makecell{Grazing incidence \\ AR plates} & $>$95$\%$ & $\sim$50$\%$ &  \makecell{High laser \\ damage threshold}  & Polarization sensitive & \cite{Takahashi04,Pronin11}\\ 
 \hline
  \makecell{Annular beam HHG \\ + pinhole} & $>$95$\%$ & $\sim$100$\%$ & Preserves polarization & \makecell{Loss of driving \\ laser energy} & \cite{Peatross94}\\ 
 \hline
   \makecell{Noncollinear HHG} & $100\%$ & $100\%$ & \makecell{Angular separation \\ of harmonics} & \makecell{Polarization sensitive} & \cite{Fomichev02,Bertrand11}\\ 
 \hline
\end{tabular}
\caption{\textbf{Methods to remove the driving fundamental beam in HHG-based XUV beamlines.} Pros and cons are defined according to the required characteristic of our beamline, described in the text.}
\label{table1}
\end{table}
\end{center}

Another elegant scheme is to drive HHG with two synchronized pulses in a non-collinear geometry~\cite{Fomichev02,Bertrand11}. Indeed, because of linear momentum conservation rules, and since both pulses contribute to the HHG process, angular separation of the harmonics (and the drivers) in the far-field can be achieved. However, the strong modulations of the laser intensity in the generating medium can induce phase mismatch and be detrimental to the overall efficiency of the process~\cite{mairesse07}. 

The last solution, which is the one that we are using, relies on spatial filtering when driving HHG using an annular beam~\cite{Peatross94,Klas18}. Generating XUV pulses using an annular beam for subsequent spatial filtering in the far-field is an old idea, dating back to the early days of HHG~\cite{Peatross94}. Indeed, while the collimated driving beam is annular, the XUV beam is Gaussian-like and propagates in the inner cone of the driving laser with lower divergence. Thus, placing a pinhole in the far-field easily allows separating the driving laser from the XUV beam. Since its initial implementation in 1994~\cite{Peatross94}, this method has extensively been used in attosecond pump-probe experiments with Ti:Sa lasers~\cite{Paul01} and was recently extended to high-repetition-rate Yb fiber sources~\cite{Klas18}.

The easiest way to generate an annular beam is through a reflection of a Gaussian beam onto a drilled mirror. In our setup we placed a drilled mirror with a hole of 4~mm diameter, after increasing the beam size with a $\times$4 telescope. The hole transmits around 20$\%$ of the laser power. The XUV will be spatially selected by placing a fixed pinhole (7~mm diameter) in front of the XUV spherical mirror and a motorized adjustable iris before the exit of the HHG chamber, allowing to reach a driving laser suppression of $\sim$10$^{-4}$ ($\sim$ 10~mW copropagating with the XUV). 

\section{Scaling of the XUV flux}~\label{sect_scaling}

In this section we describe the optimization of the XUV flux using the annular laser beam. To optimize the XUV flux, one needs to consider both the single-atom response and the macroscopic build-up of the XUV field (phase-matching). The single-atom dipole response depends on the laser peak intensity and typically scales as a power-law ($I^{3-4}$). Increasing the laser intensity thus yields a nonlinear enhancement of the single-atom response. However, the ionization of the medium induces dispersion, which influences phase-matching. To optimize the macroscopic HHG response in a gas jet, two conditions must be fulfilled~\cite{Constant99} : $l_{jet} > 3 l_{abs}$ and $l_{coh} > 5 l_{abs}$, where $l_{jet}$ is the length of the gas jet, $l_{abs}$ is the absorption length and $l_{coh}$ is the coherence length. $l_{coh}$ depends on the dispersion of the neutral atoms (positive term), the dispersion of free electrons (negative term), and the Gouy phase (negative term). More details about calculating these terms can be found in~\cite{Rothhardt14,Heyl16}. In a nutshell, they depend on the gas jet properties, the focusing geometry, as well as the ionization fraction $\eta$.

When using high laser intensity, free electrons dispersion increases. This can be compensated up to a certain point by increasing the (neutral) gas density. However, if the ionization fraction $\eta$ becomes higher than the critical ionization fraction $\eta_c$ (around $17\%$ for Ar at 21.6 eV with a 515 nm driving laser)~\cite{Paul05} the electronic dispersion cannot be compensated by increasing the gas density. Thus, a fine equilibrium between driving laser intensity and gas density is to be found to optimize the total macroscopic XUV flux. 

\begin{figure}[h]
\begin{center}
\includegraphics[width=1\textwidth, keepaspectratio=true]{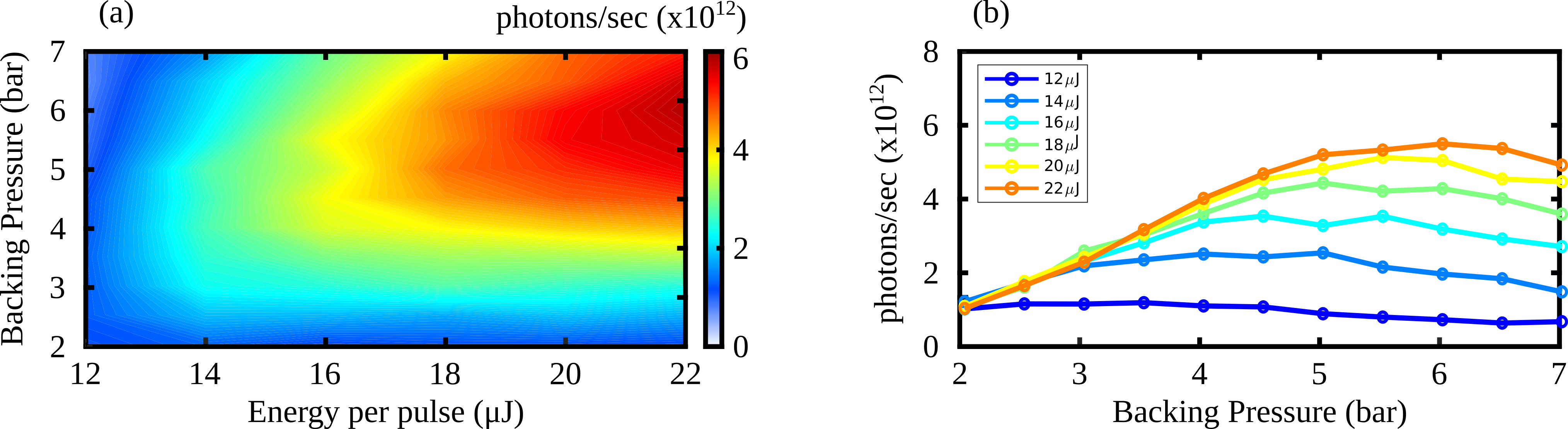}
\caption{\textbf{Scaling of XUV photon flux around 21.6~eV.} \textbf{a-b} Scaling of the flux (photons/sec) around 21.6~eV as a function of driving laser pulse energy and of the backing pressure of argon. These measurements were taken using a glass gas nozzle with a throat diameter of 50 $\mu$m.}
\label{Fig_flux}
\end{center}
\end{figure}

We measured the XUV flux (around 21.6~eV) as a function of laser pulse energy and argon backing pressure, using a gas jet with a nozzle throat diameter of 50 $\mu$m. The data are shown in Fig. \ref{Fig_flux}. To rationalize these results, we estimated $l_{coh}$ and $l_{abs}$ under our experimental conditions. Given the proximity of the laser focus to the gas jet, the argon density at the laser focus is estimated to be on the order of $40\%$ of the backing pressure~\cite{Rothhardt14}. We estimated the ionization fraction using the closed-form analytical expression for the ionization rate derived by Yudin and Ivanov~\cite{Yudin01}. Taking into account the geometric parameter of the focused laser beam, we have all the parameters needed to derive $l_{abs}$ and $l_{coh}$, and to rationalize the scaling of the XUV flux with gas pressure and laser intensity that we measured experimentally. 
For the lowest pulse energy investigated (12~$\mu$J/pulse), the XUV flux does not exhibit a strong pressure dependence. This can be understood from the fact that in this low-intensity regime, increasing the gas pressure leads to the concomitant increase of the number of emitters and a decrease of the coherent length (e.g. $l_{abs}$ = 15~$\mu$m and $l_{coh}$ = 150~$\mu$m for a backing pressure of 2 bar and to $l_{abs}$ = 5~$\mu$m and $l_{coh}$ = 22~$\mu$m for a backing pressure of 6~bar).

More interestingly, when increasing the laser intensity, we observe a general increase in the XUV flux. This general trend can be easily understood in terms of the scaling of the single-atom response with laser intensity. Moreover, as the intensity increases, the pressure at which the XUV flux maximizes shifts towards higher values. This reflects the need to compensate for the enhanced electronic dispersion by increasing the (neutral) gas density, to fulfill phase-matching conditions. We estimate that when using 22~$\mu$J/pulse, an ionization fraction of $\eta = 12\%$ is reached. At this energy per pulse, using 6 bar backing pressure yields $l_{abs}$ = 5~$\mu$m and $l_{coh}$ = 800~$\mu$m. We can thus safely claim that we are in a quasi-perfect condition for fulfilling phase-matching, according to $l_{jet} > 3 l_{abs}$ and $l_{coh} > 5 l_{abs}$ criteria.

In these optimized conditions, we measured, using XUV photodiode calibrated by the Physikalisch-Technische Bundesanstalt Berlin, a photon flux of $5.7 \times 10^{12}$ photons/s at 21.6 eV (21~$\mu$W) using 5.5~W (on target) of 515~nm at 250~kHz and an argon backing pressure of 6 bar. This corresponds to a conversion efficiency of $3.8 \times10^{-6}$. This conversion efficiency is very similar to the value recently reported by some of us~\cite{Comby19,Comby20} using the same laser source, but is reached here in a much tighter focusing regime, and using an annular driving beam. 

\section{Spectral selection around 21.6~eV}

Driving HHG using a multicycle pulse yields a train of attosecond bursts separated by half optical period with a femtosecond envelope slightly shorter than the duration of the driving laser pulse~\cite{Glover96,Schins96,Varju05,Mairesse05}. In the spectral domain, this is characterized as a comb of odd harmonics separated by twice the driver photon energy. For most photoelectron spectroscopy experiments, e.g. trARPES, a single harmonic (\textit{q}th order) must be isolated with good spectral contrast. Indeed, the presence of neighboring harmonics produces undesired replicas of the photoelectron spectrum. In trARPES, requirements regarding spectral contrast are even more critical for the \textit{q}+2 order (compared to the \textit{q}-2 order). Indeed, in such an experiment, one is often interested in monitoring the light-induced excited state population dynamics, lying within a pump photon energy above the Fermi energy and being orders of magnitude smaller than the signal coming from the occupied band at equilibrium. Any contamination from the \textit{q}+2 harmonic order would undesirably add a replica of the valence band onto this time-dependent signal, strongly hampering the signal-to-noise ratio in the region of interest. Thus, the \textit{q}/\textit{q}+2 spectral contrast is very important for time-resolved photoelectron spectroscopy. 

One obvious solution to select a single harmonics within the generated frequency comb is to use a monochromator based on a single diffraction grating. This technique is often used at synchrotron beamlines but has the fundamental drawback of generating pulse front tilt~\cite{Bor93} which hampers temporal resolution in time-resolved experiments. As mentioned before, double-grating configurations preserve the length of the optical paths of different diffracted rays, maintaining the short duration of the XUV pulse~\cite{Poletto07,Frassetto11}.  

An alternative technique for the isolation of a single harmonic can be achieved through reflections on a pair of multilayer mirrors. The multilayer mirrors can be designed to have a high reflectivity for a single harmonic. Because the reflectivity depends on the angle of incidence, this scheme enables selecting different harmonic orders by tuning the angle of incidence of the XUV beam on the multilayer mirror~\cite{fedorov20}. However, this technique has only been implemented in the soft-x-ray spectral range (e.g. around $\sim$ 100~eV), where the reflectivity of multilayer mirrors is easier to engineer. 

\begin{figure}[h]
\begin{center}
\includegraphics[width=1\textwidth, keepaspectratio=true]{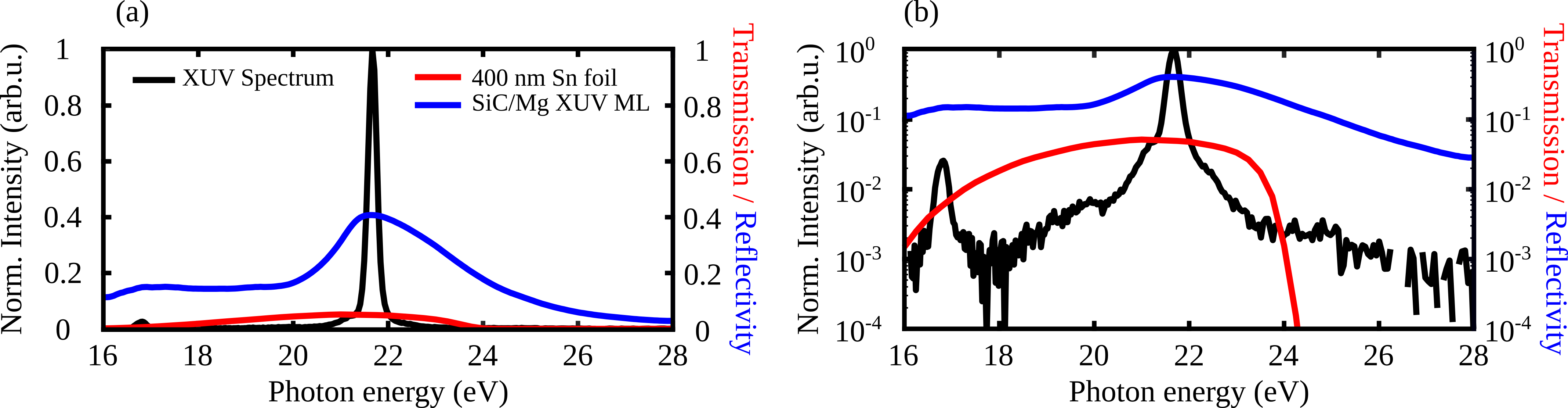}
\caption{\textbf{Monochromatization of the XUV source.}  The theoretical reflectivity of the SiC/Mg mirror (blue), the theoretical transmission of the 400~nm thick Sn foil (red) as well as the experimentally measured monochromatized XUV spectrum (black) in linear \textbf{(a)} and logarithmic \textbf{(b)} scale.}
\label{Fig_mono}
\end{center}
\end{figure}

In our case, we use a simple solution based on the combination of reflective and transmissive elements, which enable isolating a single harmonic within the comb. Indeed, the large spectral spacing (4.8~eV) between harmonics allows us to use the combination of a single reflection on a multilayer mirror (SiC/Mg, NTTAT), followed by transmission through a free-standing tin foil fixed in a gate valve. The theoretical reflectivity of the SiC/Mg mirror, the theoretical transmission of the 400~nm thick Sn foil as well as the monochromatized XUV spectrum are shown in Fig.~\ref{Fig_mono}(a)-(b) in linear and logarithmic scale, respectively.  While the harmonic 7 (16.8~eV) is two orders of magnitude less intense than the harmonic 9 (21.6~eV), we do not observe any contamination from harmonic 11 (26.4~eV), which makes our source fully compatible with the spectral contrast requirement of trARPES. It is important to note that the measured $\sim$ 230~meV spectral bandwidth (FWHM) of the harmonic 9 around 21.6~eV is determined by the poor resolution of our XUV spectrometer. As a matter of comparison, a 60~fs Fourier transform-limited pulse centered around 21.6~eV is associated with a spectral bandwidth of $\sim$ 30~meV.  The natural temporal phase of the HHG process is expected to induce a spectral broadening. This leads us to estimate that the spectral width of our source should be below 50 meV~\cite{Gauthier20}.  

\section{All-reflective XUV quarter-wave plate and polarization characterization}\label{sect_polar}

Circularly polarized XUV radiation is a very powerful tool to interrogate the properties of matter using different types of circular dichroism. For this reason, many synchrotron beamlines around the world are dedicated to static XUV circular dichroism experiments in atoms, molecules, and solids in their ground state~\cite{Derossi95,Maruyama99,Tanaka09,Nahon12,Holldack20}. One of the burgeoning goals in the community is to extend such XUV dichroism experiments to the ultrafast regime, where dynamical out-of-equilibrium properties of matter can be investigated~\cite{Willems15,Schuler20-5,Hennes21,Leveille22,Schuler22-2,Facciala22}. The ultrashort duration of XUV pulses generated using high-order harmonic generation makes this approach very appealing for pushing this goal forward. However generating circularly polarized XUV radiation using HHG is not straightforward, since the mechanism at the heart of the process, based on recolliding electron wavepackets driven by the strong laser field, has a natural preference for linearly polarized fields. Indeed, the HHG yield decays exponentially with the driving laser ellipticity ~\cite{Budil93}. Moreover, the ellipticity of the XUV radiation is similar to or smaller than the one of the driving laser (in the case of standard above-threshold non-resonant harmonic generation)~\cite{Weihe96,Antoine97}. 

In the last decade, a myriad of strategies has emerged to generate elliptically and circularly polarized HHG-based ultrafast XUV pulses. One of these approaches is based on resonant or below-threshold HHG using an elliptical driving field - in this case, a strong orthogonal component to the harmonic field is generated, yielding elliptically polarized XUV emission~\cite{Weihe96,Antoine97,Ferre15,Ferre15-2}. This technique was shown to be quite efficient to produce elliptically polarized XUV radiation in both Ar and SF$_6$ using 400~nm and 800~nm drivers. However, it was recently demonstrated that using relatively long (130~fs, as in our setup) 515~nm pulses yield XUV pulses with very small ellipticities in our spectral range of interest~\cite{Comby20}. This technique is thus not well suited in our case. 

An alternative scheme is based on the generation of two collinear phase-locked orthogonally polarized high-order harmonic sources~\cite{Azoury19}. In this approach, the polarization state of the XUV radiation is controlled by adjusting the relative delay between the two harmonic sources. This strategy is analogous to changing the thickness of a birefringent crystal. However, this solution is not only challenging to implement but also very demanding in terms of interferometric stability, which could induce difficulties in time- and angle-resolved photoelectron spectroscopy measurements which require very long acquisition times.  

Moreover, it has been demonstrated that driving HHG with non-collinear circularly polarized driving lasers yields angularly isolated XUV beams~\cite{Hickstein15}. Because of linear momentum conservation rules, angular separation of the harmonics (and the drivers) in the far-field can be achieved. While the production of both left and right circularly polarized harmonics at the same wavelength and propagating in different direction is perfectly well suited for optical spectroscopy (e.g. magnetic circular dichroism)~\cite{Hickstein15}, it would not be ideal as a source for photoemission spectroscopy. 

Another emerging strategy to generate circular XUV pulses using HHG is based on driving the process using counter-rotating bicircular bichromatic ($\omega$ and 2$\omega$) pulses~\cite{Eichmann95,Fleischer14,Kfir2015}. Indeed, the electrons which are strong-field ionized by this type of waveform can rescatter three times per laser cycle onto the parent ion, leading to the relatively efficient emission of XUV radiation. In this case, the emerging HHG spectrum appears as a succession of harmonics at 3N+1 and 3N+2 multiples of the fundamental driving frequency, while the 3N lines are forbidden, because of conservation of spin angular momentum. Indeed, since each emitted XUV photon can only carry one unit of spin angular momentum, the difference in the number of photons absorbed from the fundamental and the second harmonic needs to be $\pm$1. Due to the conservation of spin angular momentum, the 3N+1 and 3N+2 harmonics are circularly polarized and have the same helicity as the $\omega$ and 2$\omega$ fields, respectively. We have recently used this scheme and two multilayer mirrors to select a single harmonic at 34.9~eV~\cite{Comby20}. However, extending this technique to lower photon energies is very challenging, because narrowband multilayer mirrors are difficult to produce in this range.

An alternative solution, which is the one that we chose to implement here, is to introduce polarization-shaping elements~\cite{Hochst94,Vodungbo11,Willems15,Siegrist19} into the linearly polarized XUV beamline. A set of four mirrors under well-chosen incidence can play the role of an XUV quarter-waveplate, producing radiation with high degrees of ellipticity. This approach has the advantage of using the simplest configuration for HHG, i.e. driving the process using a single linearly polarized pulses. 

To generate circularly polarized XUV light, we built an all-reflective XUV quarter-wave plate based on a four SiC mirrors phase-shifter. Theoretically, the four reflections on SiC mirrors at 78$^\circ$ angle of incidence yield a phase-shift of $\lambda$/4 between s- and p- polarization components at 21.6~eV, while maintaining a high overall transmission. In contrast to standard XUV phase-shifters which rotate in the plane perpendicular to the light propagation direction under vacuum to control light polarization, our XUV quarter-wave plate is fixed. To control the polarization state of light, we rotate the linear polarization axis direction impinging on the four SiC mirrors simply by rotating the half-wave plate in front of the HHG chamber. This allows us to finely control the ratio between s- and p- polarization components, which, in addition to the fixed $\lambda$/4 phase-shift, should allow to achieve quasi-circularly polarized XUV radiation. The major advantage of this design is that switching between different light helicity does not involve any rotation of reflective optics, which often induces detrimental beam misalignment. Indeed, even recent development of rotating XUV quarter-wave plates with very good mechanical design~\cite{Yao20} have reported beam pointing shift of $\leq$ 10 $\mu$rad upon helicity swapping. Such shift would be adversarial to the measurement of CDAD in ARPES on small samples, for example. Note that we can translate the four SiC mirrors to bypass the XUV quarter-wave plate and directly use the linear polarization tunability of the XUV pulses, which can be used, for example, to measure linear (LDAD) or Fourier dichroism in photoelectron angular distributions (FDAD).  

\begin{figure}[h]
\begin{center}
\includegraphics[width=1\textwidth, keepaspectratio=true]{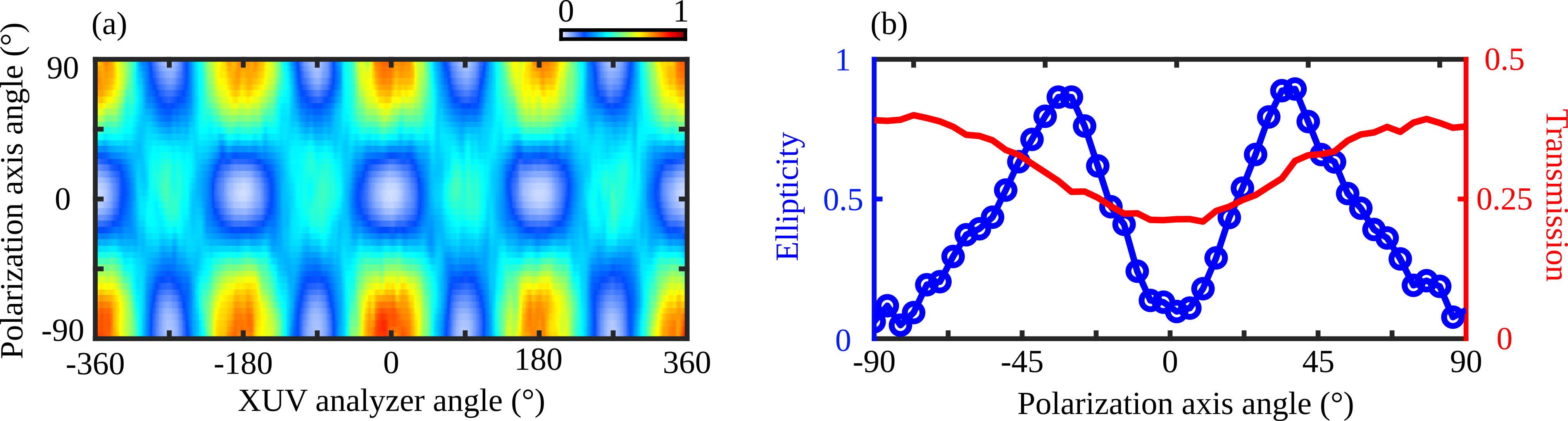}
\caption{\textbf{Characterization of the polarization state of the XUV source.} \textbf{(a)} Malus laws recording the XUV intensity while rotating the XUV analyzer angle, for different polarization axis angle impinging on the XUV quarter-wave plate. 0$^\circ$ corresponds to a p-polarization and $\pm$90$^\circ$ correspond to a s-polarization. \textbf{(b)} Ellipticity extracted using Fourier analysis for each polarization axis angle impinging on the XUV quarter-wave plate, as well as the associated measured overall transmission.}
\label{Fig_polar}
\end{center}
\end{figure}

To measure the ellipticity of the XUV for each impinging polarization axis angle, we use a Rabinovitch-type XUV analyzer~\cite{Rabinovitch65} made of four unprotected gold mirrors under 80$^\circ$ angle of incidence. This analyzer has an extinction ratio between s- and p- polarization components above 20 at 21.6~eV. It can be rotated under vacuum to provide Malus’ law curves, from which the polarization direction and ellipticity of the radiation can be extracted, under the assumption that the light is fully polarized~\cite{Antoine97}. We have to keep in mind that this assumption is not always valid and that in the end, only the measurement of a dichroic signal~\cite{Barreau18,Veyrinas16} or the use of an XUV quarter waveplate~\cite{Koide91,Nahon04,Huang18} can provide the accurate value of the ellipticity, the current method only provides upper bound values. However, while depolarization can naturally emerge from the spatial or temporal inhomogeneities of the HHG process in an elliptical laser beam, the generation of linearly polarized XUV by linearly polarized laser followed by the conversion to elliptical polarization by the reflective phase-shifter is very likely to produce a high degree of polarization.

For each impinging XUV polarization axis angle, we record Malus’ law curves by continuously rotating the analyzer by 720$^\circ$~\ref{Fig_polar}(a). As expected, the XUV signal is modulated with a period of 180$^\circ$, both with respect to the polarization axis angle and the XUV analyzer angle. In addition, we can see a $\pi$ phase-shift in the Malus’ law curves when rotating the polarization axis angle from e.g. $\pm$90$^\circ$ to 0$^\circ$. This corresponds to switching from s- to p- linearly polarized XUV radiation. In addition, the maximum XUV intensity is weaker for 0$^\circ$ polarization axis angle (p-polarization) compared to $\pm$90$^\circ$ (s-polarization), which can be explained by the s-/p- reflectivity ratio of SiC ($\sim$1.17 at 21.6eV, 78$^\circ$ angle of incidence).  At a tilted impinging polarization axis angle (e.g. around -35$^\circ$ and 35$^\circ$), the contrast of the Malus’ law curves is attenuated, reflecting the elliptical polarization state of the XUV radiation. Regarding the global transmission of the quarter-waveplate, it is 41$\%$ when using s-polarization (80$\%$ for each mirror), 21$\%$ (68$\%$ for each mirror) using p-polarization and 29$\%$ for tilt angles generating maximum ellipticities of $\sim$90$\%$. 

Using Fourier analysis of the recorded Malus laws, we extracted the ellipticity of the XUV radiation for each impinging polarization axis angle (see Fig.~\ref{Fig_polar}(b)). We can see that around $\pm$90$^\circ$ and 0$^\circ$ (impinging on the XUV quarter-wave plate using s- or p- polarization, respectively) the XUV pulses are linearly polarized. By tilting the polarization axis angle of the driving laser away from s- or p-, we can continuously tune the ellipticity introduced in the XUV beam. Around $\pm$35$^\circ$, the ellipticity maximizes to a value around 90$\%$. If the reflectivity of SiC was be the same for s- and p- polarization, we would expect the maximum ellipticity for $\pm$45$^\circ$, and symmetric behavior when going toward $\pm$90$^\circ$ or 0$^\circ$. The fact that the reflectivity is larger for s- polarization (red curve, Fig.~\ref{Fig_polar}(b)) shifts the ellipticity maximum towards smaller impinging polarization axis angles (more p- components) to have an equal contribution from s- and p- polarization components after the phase-shifter. It also breaks the symmetry when going towards $\pm$90$^\circ$ or 0$^\circ$. This simple scheme to generate quasi-circular ultrafast monochromatic XUV pulses at high-repetition-rate with fast and easy helicity switching is thus ideally suited for circular dichroism in photoemission spectroscopy.

\section{Long-term stability of the circular XUV source}\label{sect_stab}

\begin{figure}[h!]
\begin{center}
\includegraphics[width=1\textwidth, keepaspectratio=true]{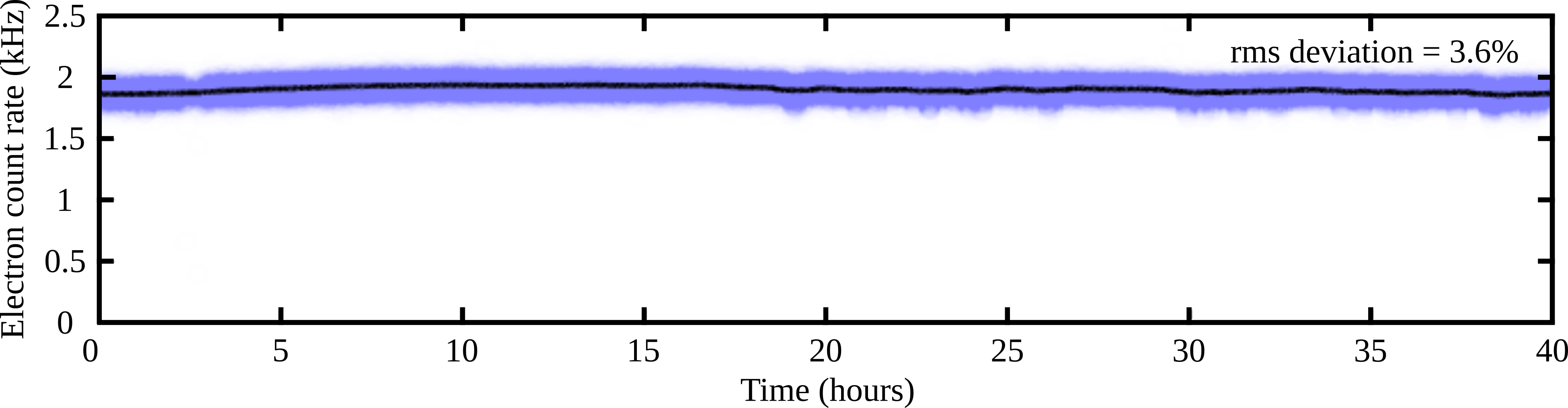}
\caption{\textbf{Long-term stability of the circular XUV source.} The blue dots represent the electron count rate as a function of time, analyzed at 1 Hz sampling rate, over 40 hours. The experimental conditions used to measure the electron count rate as a function of time are described in the manuscript. The thick black is a sliding average of the electron yield (width of 1 hour). The root means square deviation of the count rate over 40 hours is 3.6$\%$. }
\label{Fig_stab}
\end{center}
\end{figure}

To be compatible with a data acquisition time of tens of hours, necessary to generate high signal-to-noise ratio data in time- and angle-resolved photoemission spectroscopy~\cite{maklar20}, the XUV source needs to be very stable both in terms of photon flux and pointing. Moreover, measuring dichroism in photoemission requires even more statistics since the relevant physics is often encoded in small differential signals in specific regions of the highly multidimensional data. 

As a demonstration of the high stability of our circularly-polarized XUV source, we measured the single-photon ionization rate of gas-phase molecules (sampling rate 1~Hz) over 40 hours (Fig.~\ref{Fig_stab}). We sent our circularly-polarized XUV source onto a molecular beam ($\sim$ 2.3~mm FWHM, located $\sim$ 4~m from the HHG source) of $\alpha$-pinene ($I_p$ = 8.38~eV) seeded by 1.5 bar of helium and recorded the electron count rate using a COLd Target Recoil-Ion Momentum Spectrometer (COLTRIMS)~\cite{Ullrich03}. Fig~\ref{Fig_stab} shows that over 40 hours, the electron count rate is very stable (3.6$\%$ root mean square deviation). During this acquisition, the XUV helicity was changed every 10 minutes. This demonstrates the long-term stability of our setup, both in terms of photon flux and pointing, which is of capital importance for all potential applications of HHG-based XUV sources and more specifically for time-, angle-, and polarization-resolved photoemission spectroscopy.   

\section{Conclusions}\label{sect_conclu}

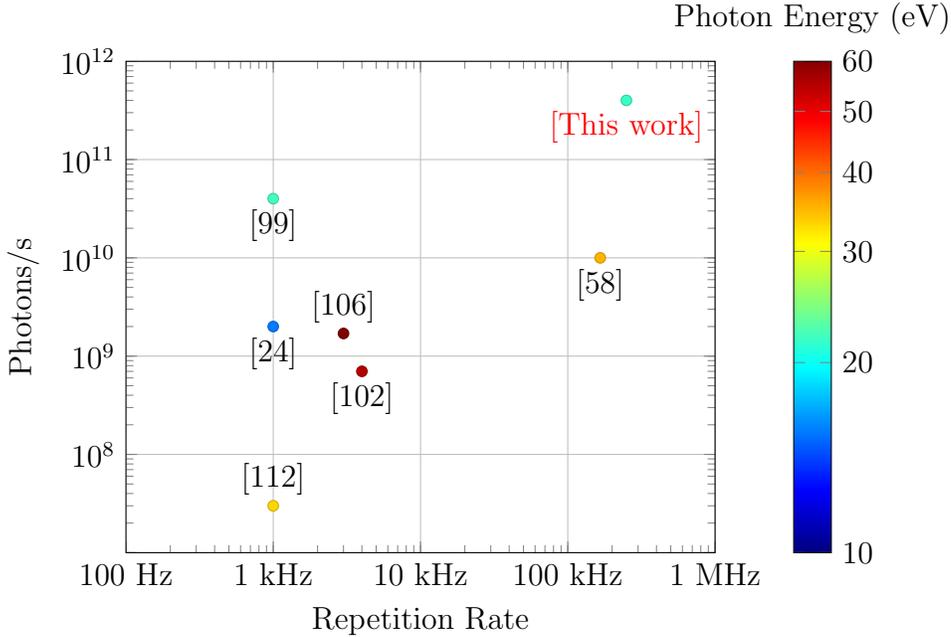
\begin{figure}[h!]
\centering
\begin{tikzpicture}
\begin{loglogaxis}[grid=major,width=0.6*\textwidth,colormap/jet,colorbar,xmin=1e2,xmax=1e6,ymin=1e7,ymax=1e12,
		point meta min=1,
		point meta max=1.77815 ,
		xlabel={Repetition Rate},
		ylabel={Photons/s},
		xtick={1e2,1e3,1e4,1e5,1e6},
		xticklabels={{100 Hz},{1 kHz}, {10 kHz}, {100 kHz},{1 MHz}},
		ytick={1e7,1e8,1e9,1e10,1e11,1e12},
		yticklabels={{},{10$^{8}$},{10$^{9}$},{10$^{10}$},{10$^{11}$}, {10$^{12}$}},
		colorbar style={title={Photon Energy (eV)},
		ytick={1., 1.30102, 1.47712, 1.6020, 1.69897, 1.77815},
		yticklabels={10,20,30,40,50,60}}
		]
    \addplot[
        scatter, mark=*, only marks, 
        scatter src=explicit,
        nodes near coords*={\Label}, 
        visualization depends on={value \thisrow{label} \as \Label}, 
        visualization depends on={value \thisrow{anchor}\as\myanchor},
   	 every node near coord/.append style={anchor=\myanchor},
    ] table [x=Hz,y=E,meta expr=log10(\thisrow{PE})] {
    
Hz	      E	      PE	label anchor
166e3	1e10	35	    \cite{Comby20} north
250e3	4e11	21.6	\textcolor{red}{[This work]} north
1e3	    4e10	22	    \cite{Hickstein15} north
1e3	    3e7	    33	    \cite{Huang18} south
1e3     2e9     15.5    \cite{Ferre15} north 
3e3     1.7e9    60      \cite{Yao20} south
4e3     7e8     55     \cite{Kfir2015}  north
 };
    
\end{loglogaxis}
\end{tikzpicture}
\caption{\textbf{State-of-the-art of quasi-circularly polarized HHG-based XUV sources.} The color of each dot represents the photon energy. The reported flux represents the flux/harmonic available on target.}
\label{bilan_flux}
\end{figure}

We reported on the development of a new polarization-tunable monochromatic ultrafast extreme ultraviolet beamline working at a repetition rate of 250~kHz. Driving HHG using an annular beam allows to spatially isolate the XUV radiation from the high power 515 nm driving laser, retaining the ability to control the polarization axis angle of the XUV. The large spectral separation between harmonics allows isolating a single harmonic (21.6~eV) from neighboring harmonics with very good spectral contrast. Our fixed XUV quarter-wave plate enables high polarization state tunability, i.e. linear polarization axis tunability as well as quasi-circularly polarized light -- up to ellipticities of 90$\%$. 

Our work is, to the best of our knowledge, the first beamline combining bright (5.7$\times$10$^{12}$ photons/s at 21.6~eV --  up to 4$\times$10$^{11}$ photons/s on target, using a 200~nm thick Sn foil) ultrafast high-repetition-rate XUV source with polarization tunability. In the past few years, important efforts have been made to optimize the photon flux of linearly polarized HHG sources~\cite{Comby19,Klas21}. Given the importance of CD experiments in various fields, we believe that such efforts also have to be conducted for circular XUV sources. To synthesize the current state-of-the-art, Fig.~\ref{bilan_flux} presents a the photon flux and photon energy of the reported quasi-circularly polarized HHG-based XUV sources as a function of their repetition rate. The source presented in this paper clearly opens new opportunities for circular-polarized photon-hungry experiments such as multi-modal dichroism in time- and angle-resolved photoemission spectroscopy from gases and solids.

\section{Acknowledgement}

We thank R. Bouillaud, F. Blais, R. Delos, N. Fedorov, and L. Merzeau for technical assistance. We acknowledge financial support from CNRS, Université de Bordeaux, Quantum Matter Bordeaux and the Région Nouvelle Aquitaine (RECHIRAM). This project has received funding from the European Research Council (ERC) under the European Union’s Horizon 2020 research and innovation program No. 682978—EXCITERS.

\section{Bibliography}

\bibliography{XUV_Beamline}
\bibliographystyle{iopart-num}

\end{document}